# Dynamic Multi-Arm Bandit Game Based Multi-Agents Spectrum Sharing Strategy Design

Jingyang Lu, Lun Li, Dan Shen, Genshe Chen, Bin Jia
Intelligent Fusion Technology, Inc.,
20271 Goldenrod Ln, Germantown, MD, 20876
{Jingyang.lu, lun.li, dshen, gchen, bin.jia}
@intfusiontech.com

Erik Blasch, Khanh Pham
Air Force Research Laboratory
Rome, NY and Kirtland, AFB, NM
{erik.blasch.1, khanh.pham}@us.af.mil

*Abstract*— **For a wireless avionics communication system, a Multi-arm bandit game is mathematically formulated, which includes channel states, strategies, and rewards. The simple case includes only two agents sharing the spectrum which is fully studied in terms of maximizing the cumulative reward over a finite time horizon. An upper confidence bound (UCB) algorithm is used to achieve the optimal solutions for the stochastic Multi-arm bandit (MAB) problem. Also, the MAB problem can also be solved from the Markov game framework perspective. Meanwhile, Thompson sampling (TS) is also used as benchmark to evaluate the proposed approach performance. Numerical results are also provided regarding minimizing the expectation of the regret and choosing the best parameter for the upper confidence bound.**

*Keywords—Multi-arm Bandit Game;Cognitive Raido Network; Dynamic Spectrum Access*

## I. INTRODUCTION

Avionics systems are dependent on communication capabilities for navigation and control [1][2]. A key element of future unmanned aerial systems (UAS) would be wireless communications. However, the many possible UAS would be sharing the available spectrum for navigation and control.

A wireless spectrum, regarded as a limited resource, has been investigated to increase the utility efficiency [1]. A Cognitive radio (CR) has been proposed to automatically adapt the communication system parameter to overcome the conflict between the great demand for spectrum and large amount of spectrum left available by Joseph Mitola III [3]. In a cognitive radio network (CRN), spectrum sensing provides the basis for the communication control center to dynamically allocate the spectrum resources without bringing harmful interference to other users [4][5].

Recently, the spectrum allocation problem has been studied from the physical (PHY), medium access control (MAC), and network layers using different approaches such as communication theory, signal processing, graph theory, machine learning, and game theory; all of which involve computational complexity and communication overhead [6]. These approaches are advancing capabilities for avionics systems. Thus, a key problem arises as how to accommodate the different parts of communication system to balance the system spectrum utility and computation complexity constrained by the limited resources.

In this paper, a new type of game is formulated to design the strategy that each communication node can dynamically select a candidate spectrum to transmit signal efficiently with the smallest accumulative regret.

In the *Multi-arm Bandit (MAB) game*, which is originally proposed in [7], a gambler has to choose one of $K$ machines to play. Each time, the gambler pulls the arm of one slot machine and receives a reward or payoff. The purpose of the game is to maximize the gambler's accumulative return or equivalently, the accumulative regret. The problem is a typical example of trade-off between the exploration and exploitation. If the player myopically focuses on the slot machine he thinks is the best, he may miss the actually best machine. On the other hand, if he spends most of time trying different slot machines, he may fail to play the best option enough often to gain an optimal reward. The traditional Multi-arm bandit game mostly depends on the assumptions about the statistics of the slot machine.

In [8], a new type of Multi-arm bandit game is investigated, in which an adversary instead of a well behaved stochastic process, has complete control over the payoffs. It is proved that the proposed algorithm can achieve the best payoff arm at the rate of $O(T^{-\frac{1}{2}})$ in a sequence of $T$ plays. Considering the high computational complexity of solving stochastic dynamic games as the number of agents grow, the proposed *mean-field approximation* can be dramatically reduced. Also, a performance bound is derived to evaluate the approximation performance [9].

Considering the computability and plausibility limitation of the Markov perfect equilibrium, an approximation methodology also called *mean field equilibrium* is considered where agents optimize only with respect to the other players' average estimate, which is reasonable because it is impossible for each player to keep knowledge of other players all the time. The necessary condition for the existence of a mean field equilibrium in such games is derived and investigated [10]. The Multi-arm bandit game is a type of sequential optimization problem, where in successive trials, an agent pulls a random arm from a given set of arms of a certain size to receive corresponding reward from unknown priori. The agent can adjust his/her strategies by only observing his/her reward history. There might possibly exist a gap between the ideal maximum reward and actual reward because of the information shortage. Based on the basic bandit problem which involves only one agent, in this paper, the problem is generalized to multiple agents, where each agent's decision will affect the other agents' reward.

Multi-arm bandit games come in two categories: stochastic and adversarial. In *stochastic case*, it is supposed that the players' action doesn't change each bandit's reward probability distribution. While in *adversarial case*, based on the agent's actions, each bandit will adjust its strategies minimizing the agent's reward on the other side. Much research has been conducted related to the jamming effect and detection in communication systems. In[11][12], a stochastic game is characterized in the threat predication and situation awareness. Based on [11], a game theoretic situation awareness and impact assessment approach is further extended for cyber network defense to consider the change of threat intent during cyber conflict [13].

In [14][15], it is assumed that the system center is unware of the existence of the jammer, where the effect of the jamming signal is studied from the trace and determinant perspective. It is common to use Chi-Square detection using the determinant for jamming detection. It is shown that in [16] that the adversary can attack the system without being detected via taking advantage of the subspace of the measurement matrix. In [17], the jammer detection is further studied in a power system, where a data frame attack is optimally designed as a quadratically constrained quadratic program. It shows that only a half critical set of measurements are needed in order to make the system unobservable.

If some prior information is accessible, Bayesian detection can be utilized for jamming detection. In [18], an improved Bayesian detection is designed that can minimize the system's estimation error instead of minimizing the detection error. A minimum mean square error estimator is used as a benchmark for performance analysis. Also, the jamming effect is also studied from game perspective [19][20], where stochastic and two-person zero-sum games are designed to improve the threat detection. Also, since transponder designs have adopted more powerful onboard processing and multiple antennas to enhance the communication quality and robustness [21][22] that can be controlled for jamming mitigation.

Overall, the contribution of this paper is summarized as follows: firstly, the Multi-arm bandit game is mathematically formulated, which includes the channel states, player strategy, and payoff reward. The simple case involves only two agents sharing the spectrum. The two-player spectrum sharing case is first fully studied in terms of maximizing the cumulative reward over a finite time horizon. An upper confidence bound (UCB) algorithm is used to achieve the optimal solutions for the stochastic MAB problems. Also, the problem can also be solved from the Markov game framework perspective. Meanwhile, Thompson sampling (TS) is also used as benchmark to evaluate the proposed approach performance.

The rest of paper is organized as follows, Section II generally describes the spectrum allocation problem and the system model. Two approaches including the UCB and the Thompson sampling are mathematically described for comparison to the MAB approach. Numerical results are shown in Section III as applied to an avionics communication system for cyber protection. Section IV summarizes the paper.

## II. SYSTEM MODEL

### A. Cognitive Radio Network

Regarding the communication limitation for avionics systems, the spectrum is divided by time and frequency. The player can choose the frequency $f_i$ with the bandwidth $B_i$ to transmit the signal information or conduct jamming activity. For the *cooperating case*, it is important for each player to sense the vacuum spectrum hole to guarantee the signal transmission performance with low probability of interference. For the *adversary case*, the player's objective is to choose the candidate spectrum hole to transmit the signal with a low probability of detection.

### B. Multi-arm Bandit (MAB) Formulation

It is supposed that a game is played among $n$ players. A stochastic game defined as $\Xi = (\chi, \mathbf{A}, \mathbf{P}, \Theta, \alpha)$.

*Time:* The game is played in discrete time, indexed by time $t = 1,2,3,\cdots$, as shown in Figure 1.

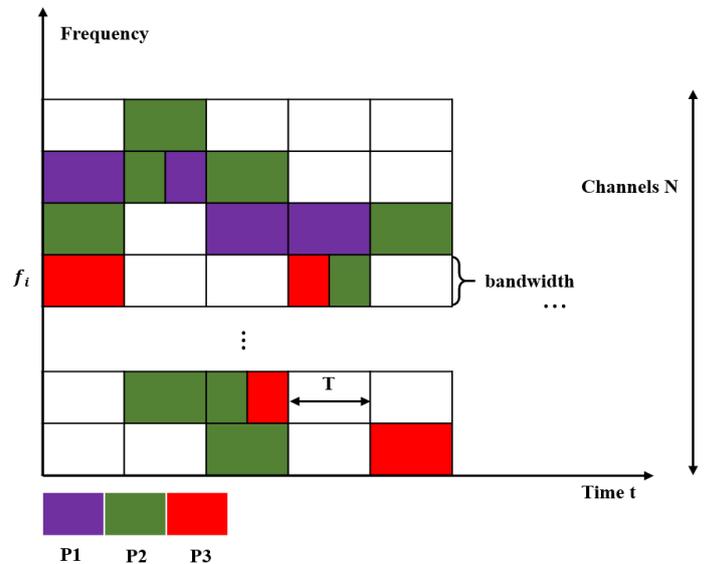

Figure 1: System Model

*State:* For each player $i$, the state at time $t$ is denoted as $x_{i,t} \in \chi$, where $\chi \in \mathbf{R}$ is compact. $\mathbf{x}_{-i,t}$ is used to denote the state of all the players except player $i$ at the time $t$.

*Strategy/Action:* The action taken by each player $i$ at time $t$ is denoted as $a_{i,t}$. The feasible action set for the state $x$ is denoted as $A(x) \in \mathbf{R}$, $\mathbf{A} = \cup_{x \in \chi} A(x)$. It is assumed that $A(x)$ is compact, so $\mathbf{A}$ is compact as well.

*Transition probability:* It is supposed that players evolve strategies based on the Markov process. Given that $x_{i,t} = x \in \chi$, $a_{i,t} = a \in A(x)$, $x_{-i,t} = \xi$, the next state for player $i$ can be characterized based on Borel probability measure $P(\cdot \,|x, a, \xi)$,

$$P(S|x, a, \xi) = P(x_{i,t+1} \in S | x_{i,t} = x, a_{i,t} = a, x_{-i,t} = \xi) \quad (1)$$

where Borel sets $S \in \chi$. Based on Equation (1), given the current state for all players and action $a$ for player $i$, the state $x_{i,t+1}$ is independent of all the other past states during the game.

*Payoff/Reward:* the payoff for player $i$ at the time $t$ is denoted as

$$r = \Theta(x_{i,t}, a_{i,t}, \mathbf{x}_{-i,t}) \quad (2)$$

Based on different types of payoffs, the Multi-arm bandit (MABs) game can be categorized as stochastic MABs, adversarial MABs, and Markovian MABs.

*Discount factor:* $\alpha$ is the discount factor and the corresponding payoff for a player till time $T$ can be characterized as,

$$R_T = \sum_{t=1}^{T} \alpha^{T-j} \Theta(x_{i,t}, a_{i,t}, \mathbf{x}_{-i,t}) \quad (3)$$

It is supposed that a gambler faces $N$ slot machines trying to find a strategy that can maximize the average reward $R_t = \sum_{t=1}^{T} r_{i,t}$, $i \in (1, \ldots, n)$, where $R_t$ denotes the accumulative reward over a finite time horizon, $r_{i,t}$ denotes the reward for each time index $t$ by choosing the arm $i$. Let $r^*$ denote the maximum reward if the player's action is supposed to be best for each round, so the goal for the gambler is to minimize the expectation of accumulative regret. In this paper, the discount factor $\alpha = 1$, the problem can be characterized as

$$\max_{i_t \in (1, \ldots, n)} \sum_{t=1}^{T} r_{i_t, t} \quad (4)$$

*Regret:* After $T$ rounds, regret is defined as the difference between the sum of the collected rewards and the sum associated with an optimal strategy rewards. The regret of an action set $\mathbf{A}$ over the sequence $(r_1, \ldots, r_T)$ is given by

$$R_T(\mathbf{A}) = R_{(r_1, \ldots, r_T)}(\mathbf{A}) = \max_{i \in (1, \ldots, n)} \sum_{t=1}^{T} r_{i,t} - \sum_{t=1}^{T} r_{i_t, t} \quad (5)$$

where $r_{i_t, t} \in r^*$ is the optimal strategy for each round.

For the player, the goal is typically to minimize the regret discussed above from either expectation, or with high probability based on the way how the rewards are generated. To determine the performance, a bound is set for an algorithm from either expectation or the high probability the reward for which the decision of the draw is selected.

As for the stochastic MABs, the objective function (5) above is investigated from the expectation perspective, which can be characterized as,

$$\mathbf{E}_{r_1, \ldots, r_T}\left[\max_{i \in (1, \ldots, n)} \sum_{t=1}^{T} r_{i,t} - \sum_{t=1}^{T} r_{i_t, t}\right] \quad (6)$$

Because the $max(\cdot)$ is inside the expectation operation, the objective function is hard to solve. Usually, the pseudo-regret objective function is considered when designing the MABs algorithm,

$$R_T(\mathbf{A}) = \max_{i \in (1, \ldots, n)} E_{r_1, \ldots, r_T}\left[\sum_{t=1}^{T} r_{i,t} - \sum_{t=1}^{T} r_{i_t, t}\right] \quad (7)$$
$$= \max_{i \in (1, \ldots, n)} E_{r_1, \ldots, r_T} \sum_{t=1}^{T} r_{i,t} - E_{r_1, \ldots, r_T}$$
$$= Tu^* - \sum_{t=1}^{T} u_{i_t}$$

where $u^*$ denotes the highest mean reward among the arms $i \in (1, \ldots, n)$, and $u_i$ denotes the reward mean for arm $i$.

For the cognitive radio network, an *arm* is regarded as one channel candidate in the limited spectrum resources. For each player $i$, the strategy is a function over the time $t$. The strategy is supposed that each player makes his/her own decision, which can be characterized as,

$$\{x_{i,t}, a_{i,t}, \Phi_t\}_{i=1}^{T-1}, \hat{x}_{i,t} \to a_{i,T} \quad (8)$$

The *Upper Confidence Bound (UCB)* algorithm is often used to find the optimal solution. Let $N_{i,t}$ denote the total times that arm $i$ has been chosen for the first T trials,

$$N_{i,t} = \Sigma_{t=1}^{T} 1(i_t = i) \quad (9)$$

and $\hat{u}_{i,t}$ denotes the sample mean rewards obtained by pulling the arm $i$ for the first T trials,

$$\hat{u}_{i,t} = \frac{1}{N_{i,t}} \Sigma_{t=1}^{T} 1 \cdot x_{i,t}(i_t = i) \quad (10)$$

The UCB algorithm is shown in Algorithm 1.

Algorithm 1: Upper Confidence Bound

| 1 | Parameter $\alpha \in [0,1]$ |
|---|---|
| 2 | For $t = 1, \ldots, T$ |
| 3 | $i^* = argmax_{i \in (1, \ldots, n)}(\hat{u}_{i,t} + \sqrt{\frac{\alpha \ln t}{2N_{i,t-1}}})$ |
| 4 | Set $N_{i,t} = N_{i,t-1} + 1$ if $i_t = i$ |

| | Else $N_{i,t} = N_{i,t-1}$ |
|---|---|
| 5 | Receive new reward $r_{i^*,t}$ |

*C. Thompson sampling*

As for the Thompson sampling, it is assumed that the player has some prior knowledge of the posterior distribution of the reward for each arm. For each arm, the player starts from a prior information on the parameters of the distribution of reward for each arm *i*. The posterior distribution of the reward for each arm is updated by taking advantage of the received observation. The best arm is selected based on the updated posterior probability. In this paper, the Bernoulli Multi-arm Bandit game is studied. $Beta(1,1)$ is usually selected as the prior information. The Thompson Sampling algorithm is summarized in Algorithm 2.

Algorithm 2: Thompson sampling

| 1 | Initialization: scale parameter $\theta_{i,1} \sim Beta(1,1)$, $H_{i,t} = 0$, $M_{i,t} = 0$ |
|---|---|
| 2 | t = t+1 |
| 3 | For each $i \in (1, \dots n)$, $\theta_{i,t} \sim Beta(H_{i,t-1} + 1, M_{i,t-1} + 1)$ |
| 4 | Choose $i_t = argmax_i(\theta_{i,t})$ |
| 5 | Go to step 2 |

### III. NUMERICAL RESULT

In this section, the *Upper Confidence Bound (UCB)* algorithm is implemented in selecting the available spectrum block to transmit the signal. It is supposed that the operator is facing four spectrum candidates regarded as four arms denoted as $A_1, A_2, A_3, A_4$, which follow one of the four distributions:

1) Bernoulli distribution $Bernoulli(1,0.5)$,
2) $Beta(4,12)$,
3) Exponential distribution with $\lambda = 9$, or
4) Finite elements ([0.25 0.5 0.75 1]) with probability ([0.3 0.3 0.3 0.1]).

The corresponding means derived based on the distribution discussed above are 0.5, 0.25, 0.11, and 0.55.

In the *first simulation*, the UCB algorithm is used to maintain the confidence intervals for the various mean rewards for each arm. For each given trial, the UCB algorithm chooses the arm with the highest upper confidence bound up to the current time *t*. In our simulation, $\alpha$ is set to be in the range(0.1,1), and the step size is 0.14. The best $\alpha$ is selected based on Equation (3). The estimation of reward mean is updated in each iteration through $\hat{u}_{i,t} + \sqrt{\frac{\alpha \ln t}{2N_{i,t-1}}}), i \in (1, \dots, n)$. The bench mark used in the simulation is that the player just chooses the arm of the highest sample mean, which is equivalently to $\alpha = 0$. The reward average and the regret average considering the different $\alpha$ values are shown in Figure 2 and Figure 4. Figure 4 shows that the UCB algorithm outperforms the naïve algorithm in terms of average reward. The UCB algorithm with $\alpha = 0.0464$ achieves the most average reward.

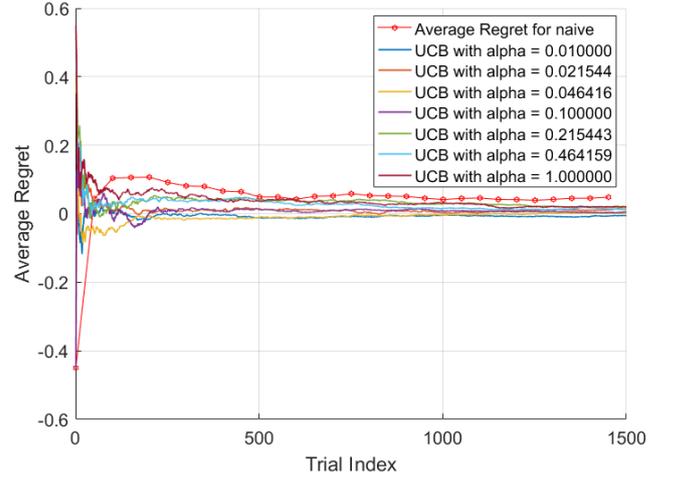

Figure 2 Average regret via UCB with different $\alpha$ value

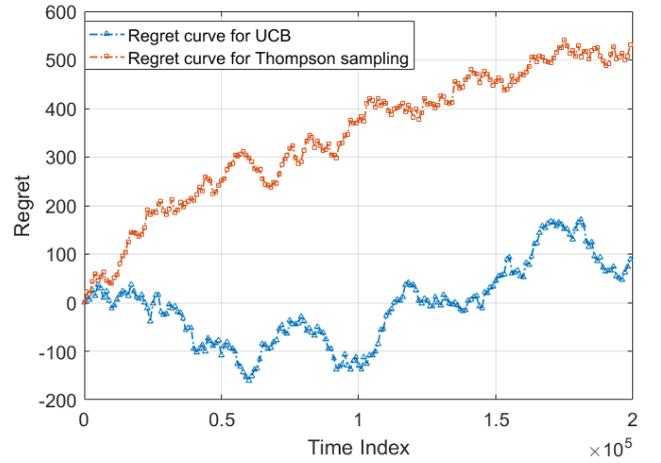

Figure 3: System performance between UCB and Thompson

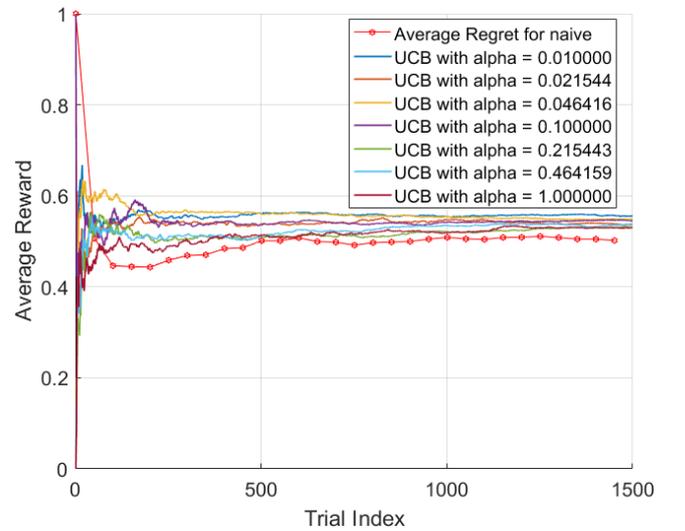

Figure 4: Average reward via UCB with different $\alpha$ value

In the *second simulation*, there are four spectrum candidates satisfying the Bernoulli distribution with mean 0.20, 0.23, 0.25, and 0.21. The optimal $\alpha$ is achieved based on simulation from the range(0.1, 1) set the same as Simulation 1. As for the Thompson sampling methods, the scale parameter is $\theta_{i,1}$ $\theta_{i,1} \sim Beta(1,1)$. From the simulation shown in Figure 3, a 200000 times Monte Carlo run is conducted in the simulation. It is shown that the UCB with the optimal $\alpha$ outperforms the Thompson sampling algorithm. The average reward between UCB and Thompson sampling is shown in Figure 5. In general, it is not practical to assume that the reward for certain arm satisfies either a Bernoulli or a Gaussian distribution, because Thompson sampling is not a Bayesian posterior sampling algorithm for general stochastic Multi-arm bandit games. However, the Thompson sampling is an online algorithm that provides a baseline for comparison. From Figure 5, it is shown that both algorithms can achieve comparatively good system performance.

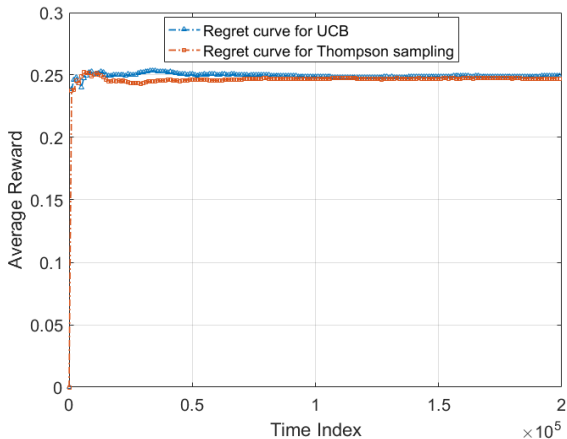

Figure 5: Average reward between UCB and Thompson sampling

## IV. CONCLUSION

In this paper, the Multi-arm bandit game is applied in the dynamically spectrum selection for avionics communications. Firstly, the Multi-arm bandit mathematically formulated, which includes channel activities, player strategy, and payoff reward. The simple case where only two agents get involved in sharing the spectrum is first fully studied in terms of maximizing the cumulative reward over a finite time horizon. The upper confidence bound (UCB) is used to achieve the optimal solutions for the stochastic MAB problems. Also, the problem can also be solved from the Markov game framework perspective. Meanwhile, Thompson sampling (TS) is also used as benchmark to evaluate the proposed approach performance.

Future work includes applying the methods for a space communications scenario [23] and for coordination of UAVs [24]. These testbeds would incorporate the methods to provide robust communication in an adversarial environment.